\documentclass[numberedappendix]{emulateapj}
\usepackage{natbib}
\usepackage{graphicx}
\usepackage{dcolumn}
\usepackage{bm}
\usepackage{amsmath}
\usepackage{amstext}
\usepackage{pstricks}
\usepackage{color}
\renewcommand{\baselinestretch}{1.18}

\newcommand\bp{\begin{figure}}
\newcommand\ep{\end{figure}}
\newcommand\bpm{\begin{figure*}}
\newcommand\epm{\end{figure*}}
\newcommand\reffig[1]{Figure \ref{fig:#1}}

\newcommand\Refsec[1]{Section \ref{sec:#1}}

\newcommand{\HI}{\mbox{H\scriptsize I}}

\def\sgras{Sgr~A$^{*}$}

\newcommand{\kevflux}{\rm \, keV\,cm^{-2} s^{-1} sr^{-1}}

\newcommand{\chandra}{\mbox{\it Chandra}}	    %

\newcommand{\be}{\begin{equation}}
\newcommand{\ee}{\end{equation}}
\newcommand{\bea}{\begin{eqnarray}}
\newcommand{\eea}{\end{eqnarray}}

\newcommand{\gev}{\rm \, GeV}

\newcommand{\degree}{^\circ}

\newcommand{\Fermi}{\emph{Fermi}}
\newcommand{\WMAP}{\emph{WMAP}}
\newcommand{\fb}{\emph{Fermi}\,bubbles\,}

\renewcommand\refname{References and Notes}

\begin{document}

\title{Evidence for Gamma-ray Jets in the Milky Way}

\author{Meng Su\altaffilmark{1,3}, Douglas P. Finkbeiner\altaffilmark{1,2}}

\altaffiltext{1}{ 
  Institute for Theory and Computation,
  Harvard-Smithsonian Center for Astrophysics, 
  60 Garden Street, MS-51, Cambridge, MA 02138 USA } 

\altaffiltext{2}{ 
  Physics Department, 
  Harvard University, 
  Cambridge, MA 02138 USA }
\altaffiltext{3}{mengsu@cfa.harvard.edu}








\begin{abstract}
Although accretion onto supermassive black holes in other
galaxies is seen to produce powerful jets in X-ray and
radio, no convincing detection has ever been made of a
kpc-scale jet in the Milky Way.  The recently discovered
pair of 10 kpc tall gamma-ray bubbles in our Galaxy may be
signs of earlier jet activity from the central black hole.
In this paper, we identify a gamma-ray cocoon feature in the
southern bubble, a jet-like feature along the cocoon's axis
of symmetry, and another directly opposite the Galactic
center in the north.  Both the cocoon and jet-like feature
have a hard spectrum with spectral index $\sim -2$ from 1 to 100
GeV, with a cocoon total luminosity of $(5.5\pm0.45)\times
10^{35}$ and luminosity of the jet-like feature of
$(1.8\pm0.35)\times10^{35}$ erg/s at $1-100$ GeV.  If
confirmed, these jets are the first resolved gamma-ray jets
ever seen.
\end{abstract}


\section{Introduction}
\label{sec:introduction}

Supermassive black holes (SMBHs) of 10$^{6}$ to 10$^{10}$
solar masses are believed to lie at the center of most
galaxies and are fed by accretion of ambient gas and stars.
Accretion-powered jets have been observed at various
astronomical scales ranging from active galactic nuclei
(AGN) \citep{Bridle:1984}, especially blazars (BL Lac
objects and flat spectrum radio quasars) at the bright end,
to gamma-ray bursts and Galactic binaries (stellar mass
black holes, neutron stars and cataclysmic variables). Some
of these objects appear to have produced jets nearly
continuously for at least tens of millions years.  The
mechanism by which jets turn on and off is one of the major
puzzles in high energy astrophysics, and may be connected to
star formation \citep{Antonuccio-Delogu}. The relativistic
jets inject significant amounts of energy into the medium
within which they propagate, creating an extended,
under-dense and hot cocoon. After decades of study, we still
lack a complete understanding of the main mechanism
launching, accelerating, and collimating jets, with limited
knowledge of the energy content, the composition, and the
particle acceleration mechanisms of the jets
\citep{Blandford:1977,Blandford:1982}.

The SMBH at the center of the Milky Way (MW) is surrounded
by clusters of young stars and giant molecular clouds
\citep{Morris:1996}. Although there are indications of past
activity \citep{Sunyaev:1993}, the SMBH is currently in a
quiescent state.  Despite the abundant observational
evidence of large-scale jets in other galaxies, it was not
expected that the Milky Way's SMBH would produce such a
relativistic collimated structure, given its current
quiescence.  However, the MW must have undergone
phases of nuclear activity in the past in order for the SMBH
to grow, and it is plausible that signs of past activity are
still visible.  One might expect relics of past activity in
high energy cosmic rays (CRs) and hot gas, perhaps far off
the disk.  The sensitivity and angular resolution of the
Large Area Telescope (LAT) on board the \emph{Fermi
  Gamma-ray Space Telescope} \citep{Atwood:2009} make
possible the search for inverse Compton (IC) gamma rays from
a Galactic jet.  

In this work, we use LAT data at 0.3 GeV $< E_\gamma < 100$ GeV
to look for unexpected diffuse Galactic gamma-ray
structure. We will show evidence for a large-scale
collimated double-jet structure, which appears symmetric with
respect to the Galactic center (GC).   For the remainder of this 
paper, we refer to these
collimated jet-like structures as the ``gamma-ray jets'' or simply the
``jets''. We also argue that the southern jet has
produced a large cocoon feature, visible over a wide range
of gamma-ray energies. These jets and cocoon features might
be associated with the previously discovered \Fermi\ bubble
structures \citep{FermiBubble}, and may hold the key to
understanding their origin.

In \Refsec{data}, we describe the \emph{Fermi}-LAT data
selection and our data analysis procedure including map
making. In \Refsec{diffusemodel}, we describe our model of
the Galactic diffuse gamma-ray emission. In \Refsec{jet}, we
show that the gamma-ray maps constructed from three-year
\Fermi-LAT observations reveal evidence of large-scale
Galactic jet features along with a south cocoon
structure. We characterize the morphology of the jet/cocoon
system in some detail and employ regression template fitting
to determine the jet and cocoon spectra in
\Refsec{jetspec}. We calculate and discuss the radio
luminosity of the jet in \Refsec{radiojet} and carefully
discuss the statistical significance of the jet structure in
\Refsec{statistics}. Finally, we discuss the implications of the
presence of the Galactic jet and future observations in
\Refsec{conclusion}.






\begin{figure*}[ht]
    \begin{center}
        \includegraphics[width=0.8\textwidth]{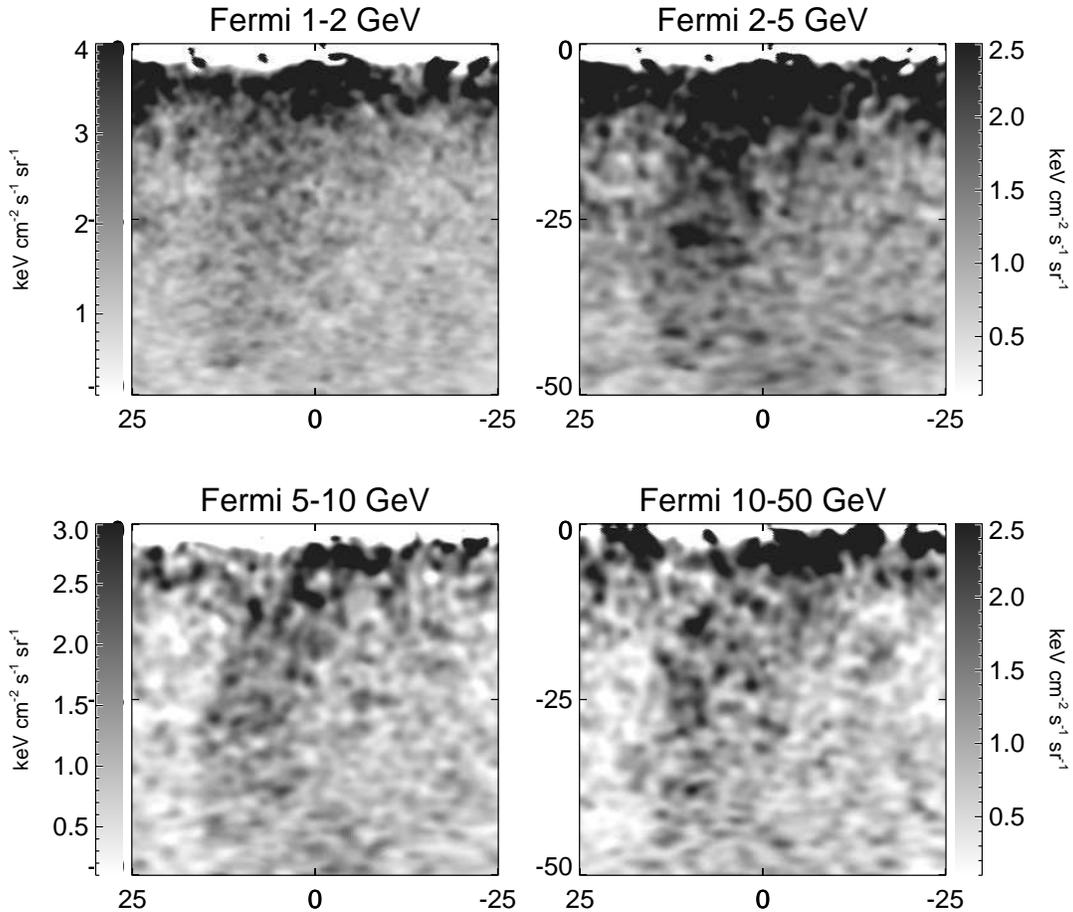}
    \end{center}
\caption{A large scale gamma-ray cocoon feature is revealed
  by template decomposition of the gamma-ray maps in energy
  range of $1-2$ GeV (\emph{upper-left}), $2-5$ GeV
  (\emph{upper-right}), $5-10$ GeV (\emph{lower-left}), and
  $10-50$ GeV (\emph{lower-right}) maps constructed from
  \Fermi-LAT three year observations.  Bright point sources
  have been masked and fainter ones subtracted.  The maps
  are smoothed by a Gaussian kernel with FWHM of
  90 arcmin. We remove the dominant diffuse Galactic signal
  from $\pi^0$ (and bremsstrahlung) gamma-rays produced by
  CR protons (electrons) interacting with the ISM using a
  dust map template constructed based on far IR data
  \citep{Schlegel:1997yv}. A smooth disk model is also
  subtracted to reveal the structure deeper into the plane.
  This model mostly removes the IC gamma rays produced by CR
  electrons interacting with the interstellar radiation
  field including CMB, infrared and optical photons. We also
  subtract a uniform template of the \Fermi\ bubbles to
  remove the corresponding gamma-ray emission. The cocoon
  feature is revealed on the east (left) side of the
  previously discovered southern \Fermi\ bubble structure
  \citep{FermiBubble}, with relatively sharp edges. }
\label{fig:cocoon}
\end{figure*}

\begin{figure*}[ht]
    \begin{center}
        \includegraphics[width=0.8\textwidth]{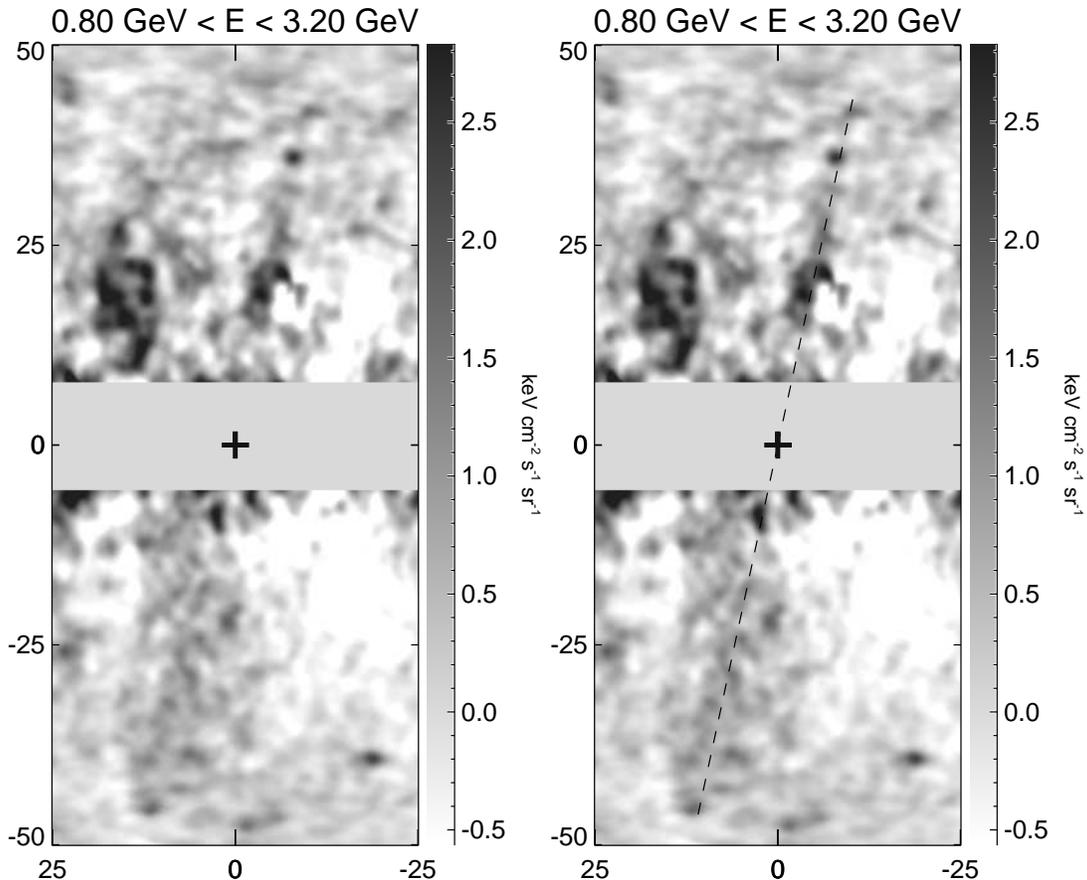}
    \end{center}
\caption{We show gamma-ray maps in Galactic coordinate
  $(-50\degree < b < 50\degree, -25\degree < \ell <
  25\degree)$ from \Fermi-LAT for photon energies $0.8-3.2$
  GeV. The Galactic center is marked with a cross sign in
  the center of the maps. Two large-scale collimated
  jet-like features (the gamma-ray ``jets'') are revealed. The
  \emph{right} panel shows the same image as the \emph{left}
  panel, but with a dashed line representing the direction
  of the suspected jet. Point sources have been subtracted
  based on the Second \Fermi-LAT catalog (2FGL)
  \citep{SecondCatalog}, and large sources, including the
  inner disk ($-5\degree < b < 7\degree$), have been
  masked. The maps are smoothed by a Gaussian kernel with
  FWHM of 2$\degree$. The \Fermi\ diffuse Galactic model
  (Pass7\_V6) has been subtracted to remove known large
  scale diffuse gamma-ray emission. }
\label{fig:fermijet}
\end{figure*}

\begin{figure*}[ht]
    \begin{center}
	\includegraphics[width=0.45\textwidth]{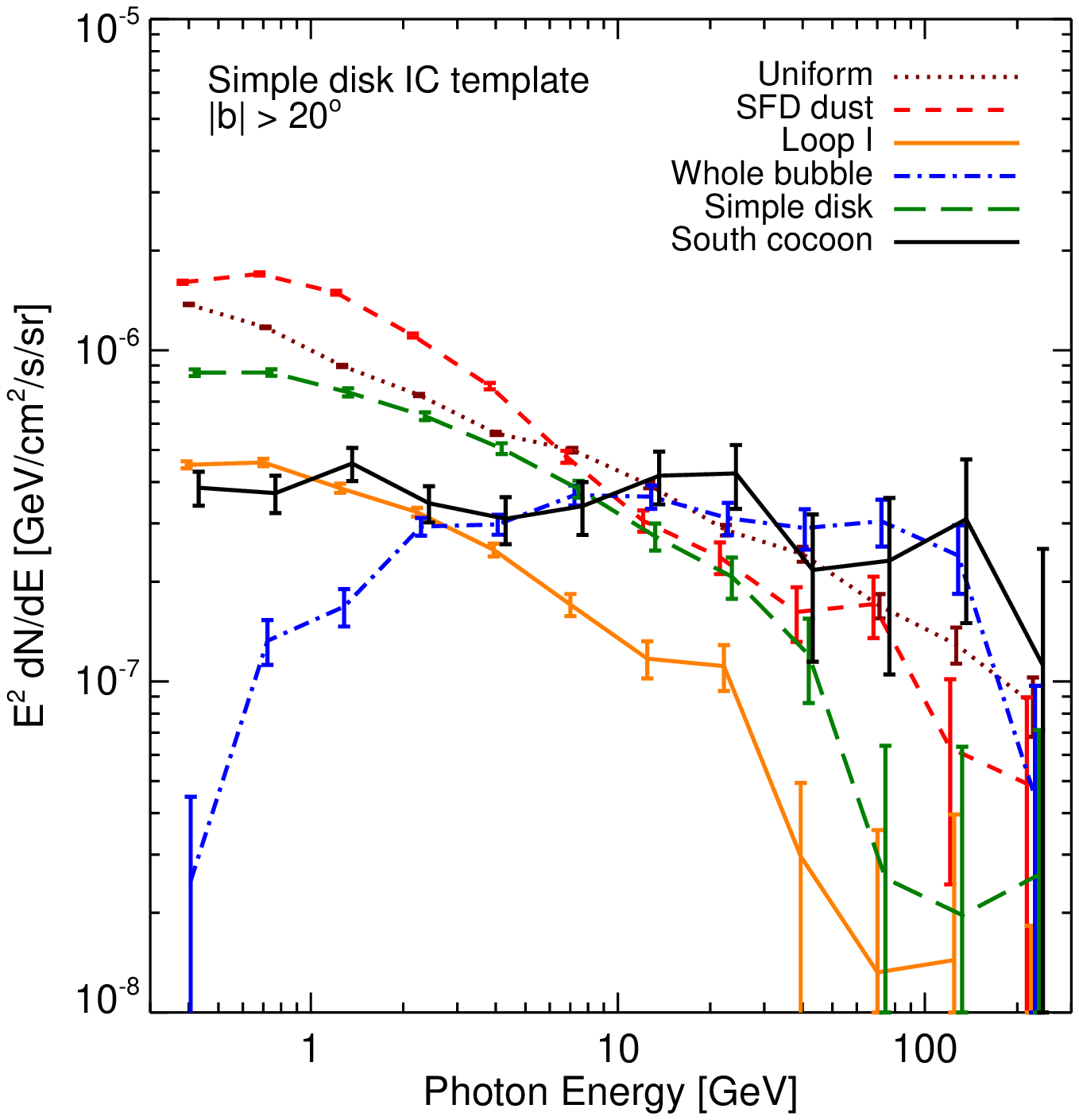}
        \includegraphics[width=0.45\textwidth]{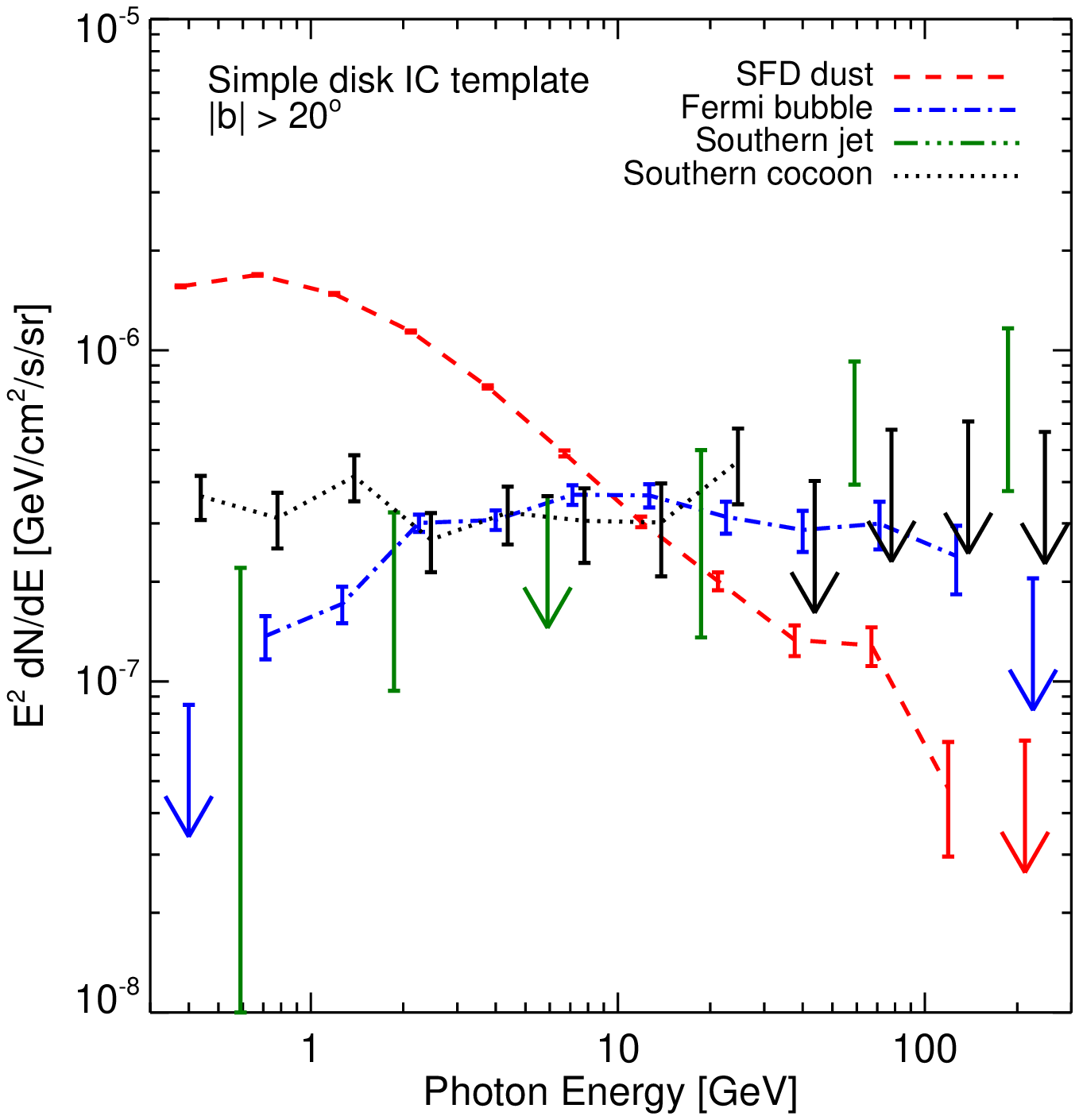}
    \end{center}
\caption{\emph{Left panel:} spectral energy
distribution of the components in our six-template fit.  The
dust-correlated spectrum (\emph{red short-dashed line}),
traces $\pi^0$ emission.  The disk-correlated emission
(\emph{green dashed}), approximately traces the soft inverse
Compton and bremsstrahlung components. The spectrum of the
uniform emission (\emph{dotted brown line}) includes the
isotropic part of the extragalactic background and
cosmic-ray contamination. The spectrum of emission
correlated with \emph{Loop I} (\emph{solid orange}) has a
spectrum similar to the disk-correlated emission.  In
contrast to these soft-spectrum components, the \Fermi\
bubble template (\emph{blue dot-dashed}) and the gamma-ray
cocoon (\emph{black solid}) have notably harder (consistent
with flat) SEDs.  Vertical bars show the marginalized 68\%
confidence range derived from the parameter covariance
matrix for the template coefficients in each energy bin.
\emph{Right panel:} same, but with one additional template
representing the gamma-ray jet.  For clarity, we only show
the spectrum of the gamma-ray jet, the cocoon and the
\Fermi\ bubbles, and compare their spectra with the softer
$\pi^0$ emission (\emph{red dashed}). }
\label{fig:spectrum}
\end{figure*}

\section{Map Construction from \emph{Fermi}-LAT Data}
\label{sec:data}

\subsection{\emph{Fermi} Data Selection}
\label{para:fermidata}

The LAT \citep{Gehrels:1999ri,
Atwood:2009} is the principal scientific instrument on
board the \emph{Fermi Gamma-ray Space Telescope}. The point-spread 
function (PSF) is about 0.8$\degree$ for 68\%
containment at 1 GeV and decreases with energy as
$r_{68}\sim E^{-0.8}$, asymptoting to $\sim$ 0.2$\degree$ at
high energy. It is designed to survey the gamma-ray sky in
the energy range from about 20 MeV to several hundreds of
GeV.

Using standard photon event selection (zenith angle $<
100\degree$, etc.) we generate full-sky maps of counts and
exposure using HEALPix, a convenient equal-area iso-latitude
full-sky pixelization widely used in the cosmic microwave background (CMB)
community.\footnote{HEALPix software and documentation can
be found at \texttt{http://healpix.jpl.nasa.gov}, and the
IDL routines used in this analysis are available as part of
the IDLUTILS product at
\texttt{http://sdss3data.lbl.gov/software/idlutils}.}
Spherical harmonic smoothing is straightforward in this
pixelization, and we smooth each map by the appropriate
kernel to obtain a Gaussian PSF of $2\degree$ FWHM. Because
the PSF of the initial map must be smaller than this, at
energies from 300 MeV to 1 GeV we use only front-converting
events, which have a smaller PSF. Above 1 GeV, we combine
front- and back-converting events.  

We make use of the \Fermi\ Pass 7 (\texttt{P7}) data
products, using the latest publicly available event
reconstruction algorithms\footnote{Details at
\texttt{http://fermi.gsfc.nasa.gov/ssc/}}.
For the purpose of studying Galactic diffuse gamma-ray
emission with minimal non-photon contamination (low
background), we select events designated
\texttt{ULTRACLEAN}\footnote{Maps produced with
\texttt{CLEAN} have about 20\% more events.  Repeating our
analysis on these maps yields similar results, with similar
error bars at $E\lesssim 10$ GeV, and slightly larger error
bars at $E \gtrsim 10$ GeV, where background is relatively
more important.} class that have the most stringent data
selection criteria, so that any diffuse features appearing
on the gamma-ray maps are not due to CR
contamination.  Photons coming from the bright limb at
Earth's horizon, dominantly produced by grazing-incidence CR
showers coming directly toward the LAT are a strong source
of contamination. We minimize this background by restricting
to events with zenith angle less than $100\degree$ as
suggested in the \Fermi\
Cicerone\footnote{\texttt{http://fermi.gsfc.nasa.gov/ssc/data/analysis/documentation/}.}.

\subsection{Map Making}
\label{para:mapmaking}

Our current gamma-ray maps (v3\_3) constructed from the
three year \Fermi\ data have greater signal/noise and
significantly lower background compared to the previously
released v2\_3 maps in~ \cite{FermiBubble}. The ``three year
maps'' exclude some short time periods, primarily while \Fermi\ passes through
the South Atlantic Anomaly.  We construct maps using the newly released photon
event list corresponding to the P7\_V6 Instrument Response Functions
(IRFs).

As in \cite{FermiBubble}, we construct maps of front-converting and
back-converting events separately, smooth to a common PSF, and then combine
them. To reveal the diffuse emission, we subtract point sources using the
Second \Fermi-LAT catalog (2FGL), which is based on 24 months of LAT
observations\footnote{\texttt{http://fermi.gsfc.nasa.gov/ssc/data/access/lat/2yr\_catalog}, the
  file we used is \texttt{gll\_psc\_v05.fit}} and the P7\_V6 event selections
and IRFs.  The PSF and effective area of the
\Fermi-LAT vary with energy, and we subtract each point source from the maps
in each energy bin, using the in-flight version of the PSF contained in the
P7\_V6 IRFs. We produce the exposure maps using the \texttt{gtexpcube} task in
the Fermi Science Tools.

In principle the PSF of each event depends on the angle
between its arrival direction and the instrument axis.  The
LAT scan strategy results in a small variation of the PSF
with position on the sky.  We use the in-flight
determination of the PSF that neglects these
details\footnote{See \texttt{http://fermi.gsfc.nasa.gov/ssc/data/analysis/documentation
    /Cicerone/Cicerone\_LAT\_IRFs/IRF\_PSF.html}}.  For the
400 brightest and 400 most variable sources, the subtraction
is noticeably imperfect, so we interpolate over the core of
the PSF after subtracting the best estimate.  We take care
to expand the mask for very bright sources (Geminga, 3C
454.3, and LAT PSR J1836+5925) and large sources (Orion and
the Magellanic Clouds). The resulting map is appropriate for the
analysis of diffuse emission at $|b| > 3\degree$.  At $|b| <
3\degree$ the maps are severely compromised by the poor
subtraction and interpolation over a large number of point
sources.  Further details of the map processing may be found
in \cite{fermihaze} and \cite{FermiBubble}.  The v3\_3 maps
used in this work are
available for download\footnote{Available at
  \texttt{http://fermi.skymaps.info}} in both FITS and jpeg formats.

\section{Diffuse Galactic Gamma-ray Emission Models}
\label{sec:diffusemodel}

At low ($\sim$1 GeV) energies, and near the Galactic plane
($|b| < 20\degree$), the gamma rays observed by \emph{Fermi}-LAT
are dominated by photons from the decay of $\pi^0$
particles, produced by the collisions of CR protons with
ambient gas and dust in the interstellar medium (ISM). Collisions of CR
\emph{electrons} with the ISM (primarily protons, but also
heavier nuclei) produce bremsstrahlung radiation.  The CR
electrons also IC scatter the interstellar
radiation field (ISRF) up to gamma-ray energies.  In order to
reveal the gamma-ray jet features, significant $\pi^0$
emission, bremsstrahlung, and IC emission from the Galactic
disk must be removed.  We take two approaches to this
foreground removal: one is to use the \Fermi\ Diffuse
Galactic Model\footnote{\texttt{http://fermi.gsfc.nasa.gov/ssc/data/}}
provided by the \Fermi\ team; the second is to employ a
linear combination of templates of known emission
mechanisms, using existing maps from multi-wavelength
observations and/or constructed geometric templates.

\subsection{\emph{Fermi} Diffuse Galactic Model}
\label{sec:fermidiffuse}


The LAT Diffuse Galactic Model is a comprehensive model of
Galactic gamma-ray emission from the ISM\footnote{Available
  from \texttt{fermi.gsfc.nasa.gov/ssc/data/}}.  The
\Fermi\ diffuse model is primarily designed as a background
template for point source analysis or investigation of
small-scale diffuse structures, and comes with a number of
caveats. However these caveats apply mainly near the
Galactic plane, and at $E > 50 \gev$.  It is nevertheless
useful for qualitatively revealing features in the diffuse
emission at high latitude. In this work, we use the version
of \Fermi\ diffuse Galactic model
\texttt{gal\_2yearp7v6\_v0.fits}.

\subsection{Simple Template-based Diffuse Galactic Model}
\label{sec:simplediffuse}

The $\pi^0$/bremsstrahlung gamma-ray intensity is
proportional to the ISM density $\times$ the CR
proton/electron density, integrated along the line of
sight. As long as the CR proton/electron spectrum and
density are approximately spatially uniform, the ISM column
density is a good tracer of the resulting gamma-ray
distribution from $\pi^0$/bremsstrahlung emission.  Because
the emission is optically thin, subtracting a simple
template of the $\pi^0$ gammas helps to reveal the gamma-ray
jets/cocoon along with the \Fermi\ bubbles, especially
towards the Galactic plane. The ISM column density is
expected to be strongly correlated with other tracers of the
ISM.  Our previous work \citep{FermiBubble} used the SFD map
of Galactic dust, based on far IR data
\citep{Schlegel:1997yv}.  The dust map has some advantages
over gas maps.  One shortcoming of using \HI\ and CO maps is
the existence of ``dark gas'' \citep{Grenier:2005}, clouds
with gamma-ray emission that do not appear in the \HI\ and
CO surveys.  These features are seen in dust maps
\citep{Schlegel:1997yv} and may simply be molecular H clouds
under-abundant in CO. For the dust map, there are no
problems with self absorption, thus no concerns about ``dark
gas,'' and the SFD dust map has sufficient spatial
resolution (SFD has a FWHM of $6'$, and LAB \HI\ is $36'$). On
the other hand, SFD contains no velocity information, so it
is impossible to break the map into Galactocentric rings as
is done with \HI\ and CO maps.  Nevertheless, it is
instructive to employ the SFD map to build a very simple
Galactic diffuse gamma-ray template. The goal is to reveal
the gamma-ray jet/cocoon structures by removing the expected
diffuse emission in a fashion that makes as few physical
assumptions as possible.


We use the SFD dust map as a template of the $\pi^0$ gamma
foreground. The correlation between \Fermi\ gamma-ray maps
and the SFD dust map is striking, and most of the ISM
emission is removed by this subtraction. To reveal the
structure deeper into the plane, a simple disk model is
subtracted \citep{FermiBubble}\footnote{The functional form
of this disk template is $(\csc|b|)-1$ in latitude and a
Gaussian ($\sigma_\ell=30\degree$) in longitude}. The disk
model mostly removes the IC gamma rays produced by cosmic
ray electrons interacting with the ISRF including CMB, infrared, and optical photons;
such electrons are believed to be mostly injected in the
Galactic disk by supernova shock acceleration before
diffusing outward. 
The two jets are symmetric with
respect to the Galactic plane and well centered on the GC,
thus they are unlikely to be local structures.



\begin{figure*}[ht]
\begin{center}
\includegraphics[width=0.8\textwidth]{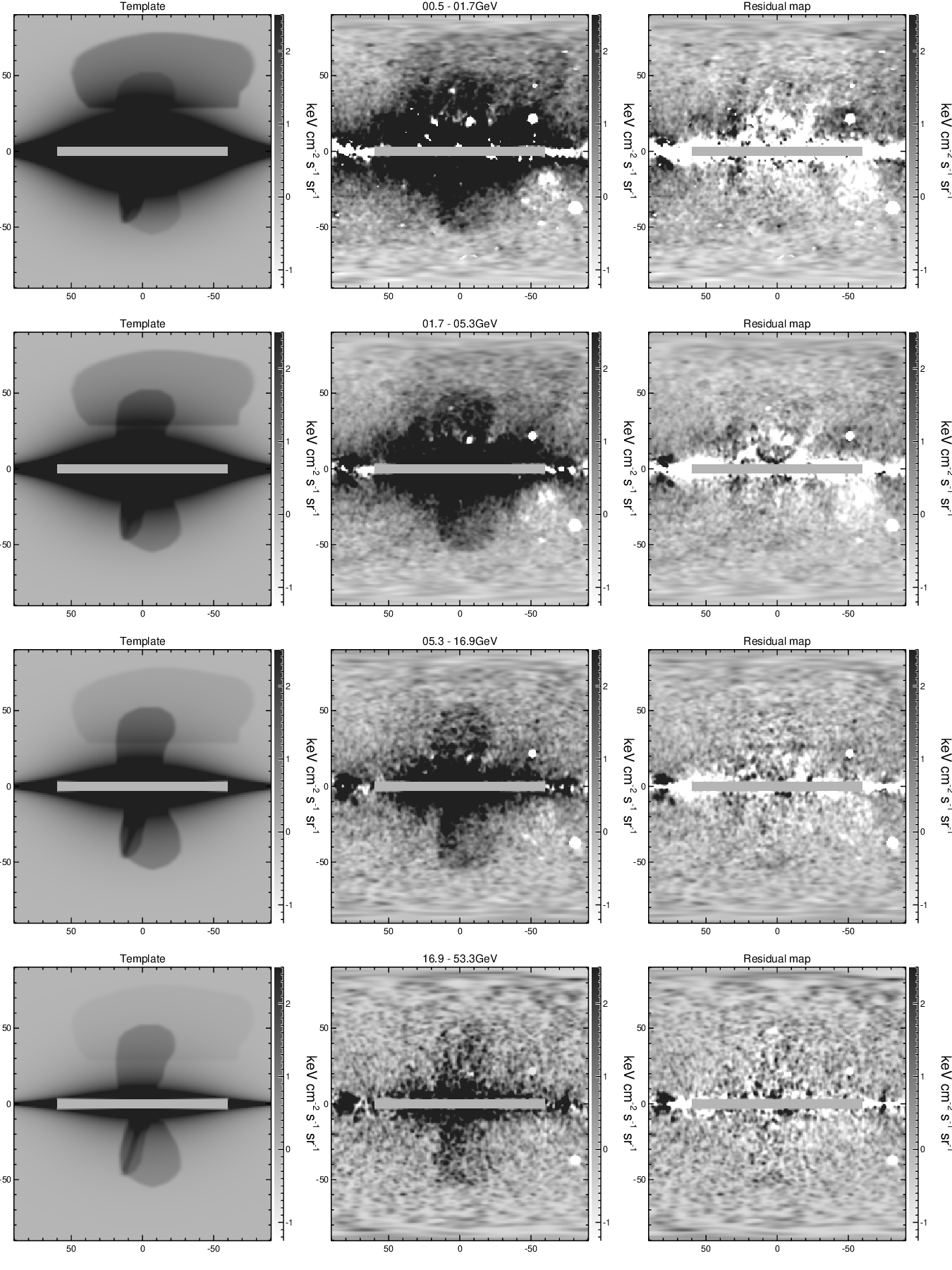}
\end{center}
\caption{This figure shows the best fit linear combination
maps corresponding to the spectra in \reffig{cocoon}.  To
increase signal/noise, larger energy bins are used.  The
\emph{left} column shows the linear combination of the disk,
\emph{Loop I}, uniform, bubble template, the south cocoon,
and the south jet template, that provide the best fit to the
\Fermi\ maps (\emph{middle column}) after subtracting the
best fit SFD dust template.  Because the $\pi^0$ emission
traced by SFD is so bright, it is subtracted from both the
models and data shown in this figure.  The difference maps
(data minus template model) are in the \emph{right}
column. The template fitting is done for the region with
$|b| > 20\degree$ to avoid contamination from the Galactic
disk. The subtraction of the model largely removes the
features seen in the \Fermi\ maps with $|b| > 20\degree$. We
have also masked the inner Galactic plane region ($|b| >
4\degree$ and $|l| > 60\degree$), which is significantly contaminated
by point sources. We use the same gray scale for
all panels. We find that both the disk IC template and
\emph{Loop I} features fade away with increasing energy, but
the jet template does not.  The oversubtraction in the
residual maps, especially in the lower energy bins, is due
to the simple disk IC model, which is not a good template
across the entire disk. However, in the fit region ($|b| >
20\degree$), the residual maps are consistent with Poisson
noise without obvious large scale features.}
\label{fig:jetmap}
\end{figure*}



\begin{figure}[ht]
\begin{center}
\includegraphics[width=0.45\textwidth]{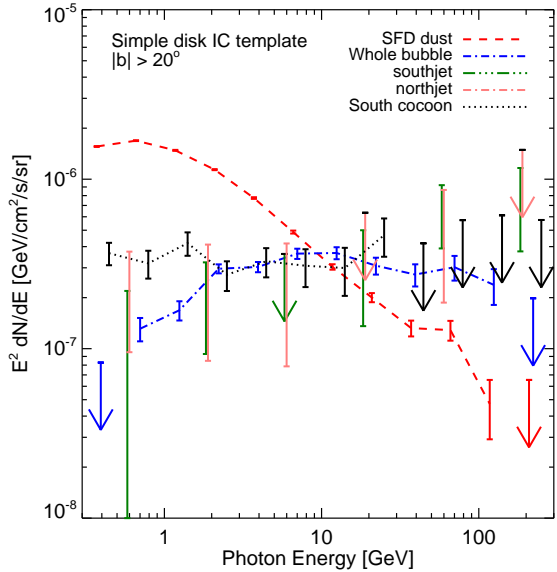}
\end{center}
\caption{This figure is the same as \reffig{spectrum}, except
    with the north jet template added to the regression. The
    north and south jet spectra are noisy, but consistent
    with each other, and consistently hard.  Note that the
    upper limit is shown for 3$\sigma$. The spectra plotted
    here are available in Tables 1 and 2.}
\label{fig:northjetspec}
\end{figure}


\begin{figure*}[ht]
\begin{center}
\includegraphics[width=0.8\textwidth]{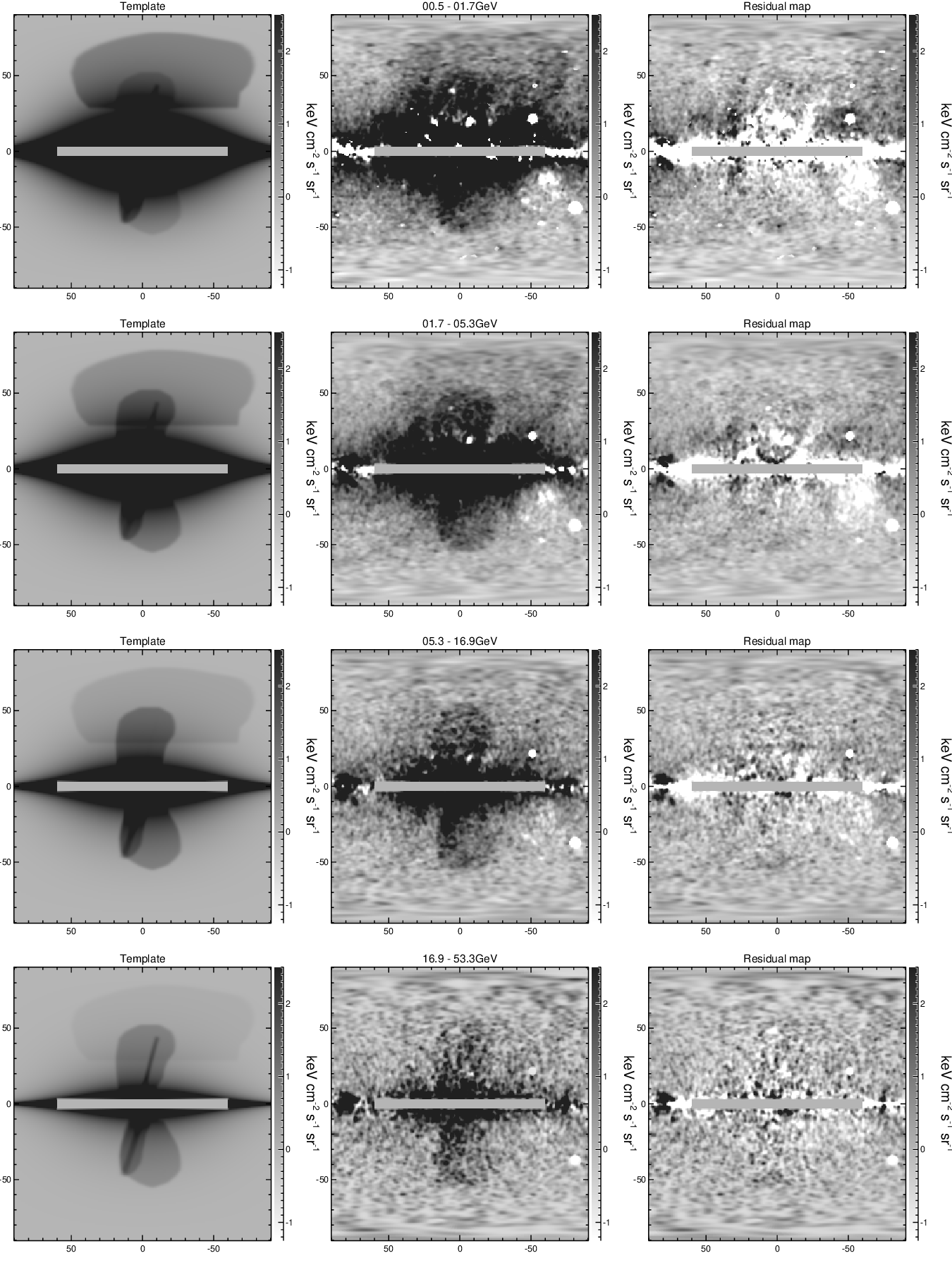}
\end{center}
\caption{Same as \reffig{jetmap}, except we include the
  north jet template. }
\label{fig:northjetmap}
\end{figure*}

\section{Gamma-ray Cocoon and Evidence of Galactic Jet}
\label{sec:jet}

The gamma-ray cocoon in the south is visible in the 3-year
LAT maps at energies from 1 to 50 GeV (\reffig{cocoon}). By inspection, the
major symmetric axis of the cocoon lines up with the
Galactic center to within $\sim 5\degree$, and the major to
minor axis ratio is about a factor of three. The gamma-ray
luminosity of the south cocoon with Galactic latitude $|b| >
20\degree$ and in the energy range $1-100$ GeV is
(5.5$\pm$0.45)$\times 10^{35}$ erg/s.  The morphology of the
gamma-ray cocoon resembles the observed radio cocoon
structures of Fanaroff-Riley (FR) type II
\citep{Fanaroff:1974} active radio galaxies (e.g. Cygnus A),
which have been found surrounding collimated large-scale
jets and may be formed by backflow of magnetized jet
plasma. How the gamma-ray cocoon system formed and retains
its tight columnar shape despite traveling from the GC for
$\sim$10 kpc is intriguing. The presence of this large scale
cocoon suggests collimated injection of high energy
particles from the inner Galaxy.

In \reffig{fermijet}, we show the $0.8-3.2$ GeV gamma-ray
map of the inner Galaxy. A pair of collimated linear
features is revealed in \reffig{fermijet}, with similar
morphology in each energy bin. There are no other apparent
large scale features in the residual maps. The gamma-ray
jets do not appear to be associated with the well-known {\it
  Loop I} structure \citep{Berkhuijsen:1973}. The north jet extends
from the GC to Galactic coordinates $(\ell, b) = (-11, 40)$,
and the south jet extends from the GC down to $(\ell, b) =
(11, -44)$.  Although the jets are faint, three lines of
evidence suggest that they are real: (1) the jets both emanate
from the GC, in nearly opposite directions; (2) the jets
extend away from the GC to about the edge of the previously
discovered \Fermi\ bubble structure \citep{FermiBubble}; and
(3) the south jet aligns with the symmetry axis of the
cocoon structure.  The morphology of the jet and cocoon
gamma-ray feature and the possible association with the
\fb\ strongly suggests the Galactic center origin of these
structures and implies recent activity toward the inner
Galaxy.

On a much smaller scale, \chandra\ X-ray observations have
revealed a faint jet-like feature at subparsec scale, pointing
toward \sgras \citep{Muno:2008}.  It is tempting to
interpret this feature as a jet of synchrotron-emitting
particles ejected from the SMBH. However, due to it's much
smaller size and $\sim$15$\degree$ misalignment with the
gamma-ray jet, there is no clear association.  Limits on the
gamma-ray jet from other wavelengths do not strongly
constrain its nature, underscoring the need for further
multi-wavelength studies.

The projected direction of the gamma-ray jets is about
$15\degree$ from the north-south axis of the Galaxy. If the
gamma-ray jets constitute the projection of double
large-scale jets symmetric to the Galactic plane and the GC,
taking the distance from the solar system to the GC
$R_\odot$ = 8.5 kpc, each projected jet is $\sim$10 kpc. The
small angle of the projected jet relative to the north-south
Galactic axis suggests that the apparent length of $\sim$10
kpc is a good approximation of the spatial scale of the
Galactic jet. The width of the jets appears to be $\lesssim
5\degree$. The total luminosity of the north and
south jet-like features is $(1.8\pm0.35)\times10^{35}$
erg/s at $1-100$ GeV (see Table 2). The jets do not align
with any plausible artifacts relating to the \emph{Fermi} orbit or
scan direction, or any other known systematics. The north
and south jets have similar integrated gamma-ray flux, so
there is no evidence that the jets are close to the line of
sight; otherwise, the approaching jet would appear
substantially Doppler brightened relative to the receding
one.



\begin{figure}[ht]
\begin{center}
\includegraphics[width=0.45\textwidth]{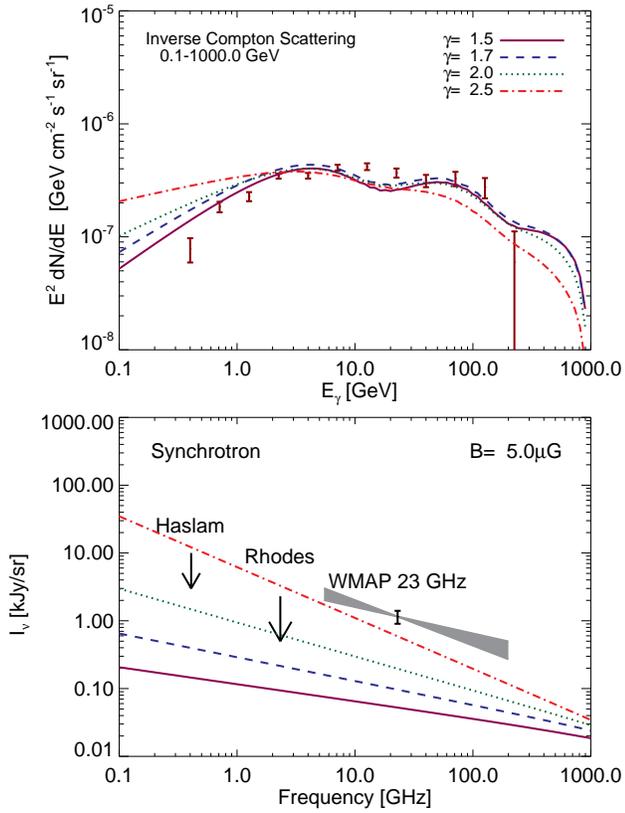}
\end{center}
\caption{The estimated spectrum of inverse Compton
gamma-rays (\emph{upper panel}) and corresponding
synchrotron radiation (\emph{lower panel}) originating from
a hard electron spectrum along a line of sight 4 kpc below
the Galactic plane (i.e. $b \approx -25\degree$). The
steady-state electron spectrum is taken to be a power law,
$dN/dE \propto E^{-\gamma}$, with index $\gamma =$ 1.5
(\emph{solid black}), 1.7 (\emph{blue dashed}), 2.0
(\emph{green dotted}), and 2.5 (\emph{red dash-dotted}) in
both the upper and lower panels. In all cases the cosmic ray
electron has a range of [0.1, 1000] GeV. The interstellar
radiation field model is taken from \texttt{GALPROP} version
50p, and the magnetic field is assumed to be 5 $\mu$G for
synchrotron calculation. The data points in the upper panels
show the south cocoon emission the same as in
\reffig{spectrum}. The arrows shows 3$\sigma$ upper limits
rather than data points with $1 \sigma$ error bars, due to
the large uncertainties in those energies. The data point in
the lower panel shows the magnitude of the \WMAP\ haze
averaged over $b=-20\degree$ to $-30\degree$ in the 23 GHz
K-band, and the gray area indicates the range of synchrotron
spectral indices allowed for the \WMAP\ haze
\citep{Dobler:2008ww}. Two arrows at lower frequency shows
the 3$\sigma$ upper limits for the cocoon radio emission
from Rhodes/HartRAO 2.326 GHz radio continuum survey
\citep{Rhodes} and Haslam 0.408 GHz map \citep{Haslam}. The
same population of cosmic ray electrons can consistently
both the radio/microwave observations and the gamma-ray
observations.}
\label{fig:radiojet}
\end{figure}

This gamma-ray jet/cocoon structure is not visible in the
\emph{ROSAT} All-Sky Survey of soft X-ray
\citep{Snowden:ROSAT}, \emph{Wilkinson Microwave Anisotropy Probe} (\WMAP) microwave maps, radio maps at
408 MHz \citep{1982A&AS...47....1H}, or any other available
radio maps.  We estimate expected radio spectra in
\reffig{radiojet} corresponding to electron spectral indices
$1.5 <\gamma < 2.5$, and find that over this range, current
full-sky radio surveys are not sensitive enough to detect
the jet. However, combining the gamma-ray energy spectrum
along with the radio limits, we find that the CR
electrons producing the jet have a hard spectrum with spectral
index $\gtrsim -2$.



We now turn to the relation between the jet/cocoon and the
\Fermi\ bubbles.  We recently discovered two giant gamma-ray
bubbles with a total luminosity from 1 to 100 GeV of
$2.0\times 10^{37}$ erg/s\footnote{We note that in
  \cite{FermiBubble}, the total gamma-ray luminosity of the
  \Fermi\ bubbles was mis-quoted as $4.0\times 10^{37}$
  erg/s. }, extending $\sim
50\degree$ above and below the GC, with a width of $\sim
40\degree$ in longitude, and found them to be spatially
correlated with a hard-spectrum microwave excess (known as
the \WMAP\ haze) \citep{Finkbeiner:2003im} and large-scale
X-ray features \citep{FermiBubble}. Galactic shock waves
produced by energetic explosions at the Galactic nucleus or
by a high rate of supernova explosions in the nuclear disk will
be channeled by the decreasing ambient gas density in
directions perpendicular to the Galactic plane. 

The gamma-ray emission associated with these bubbles has a
significantly harder spectrum ($dN/dE \sim E^{-2}$) than the
IC emission from electrons in the Galactic disk, or the
gamma-rays produced by decay of $\pi^0$ from proton-ISM
collisions. The \Fermi\ bubbles are likely formed during an
active phase in the Galactic center $\sim 10^6 - 10^7$ years
ago with jets ejected from the central supermassive black
hole. The bubble region might consist of decelerated jet
material, radiating isotropically.  Observational data and
numerical simulations indicate that the energy required to
form the bubbles is on the order of $10^{55} - 10^{58}$ erg
\citep{FermiBubble, Guo:2011, Guoetal:2011}. Even though the current jet
luminosity is relatively faint, the discovery of the
jet/cocoon system generally supports the AGN hypothesis for
the origin of the bubbles.

\begin{figure}[ht]
\begin{center}
\includegraphics[width=0.45\textwidth]{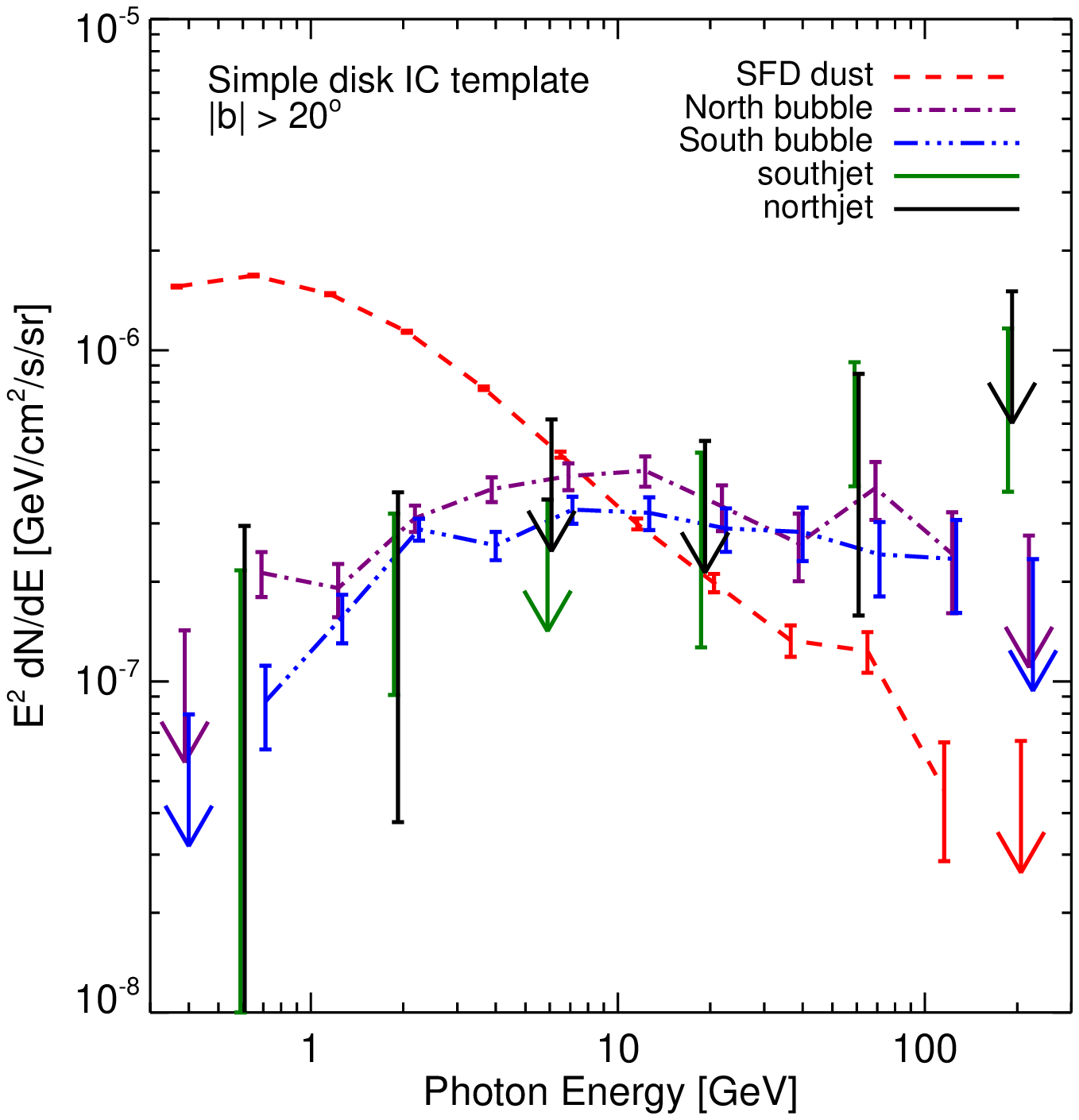}
\end{center}
\caption{Same as \reffig{cocoon}, except we add the north jet
template and split the \Fermi\ bubble into north and south
bubbles separately.  The south cocoon template is
included in the fit. The north and south jet spectra are
consistently hard. The fitting results are not significantly affected by the
bubble template splitting.  }
\label{fig:bubblesplitspec}
\end{figure}


\begin{figure*}[ht]
\begin{center}
\includegraphics[width=0.8\textwidth]{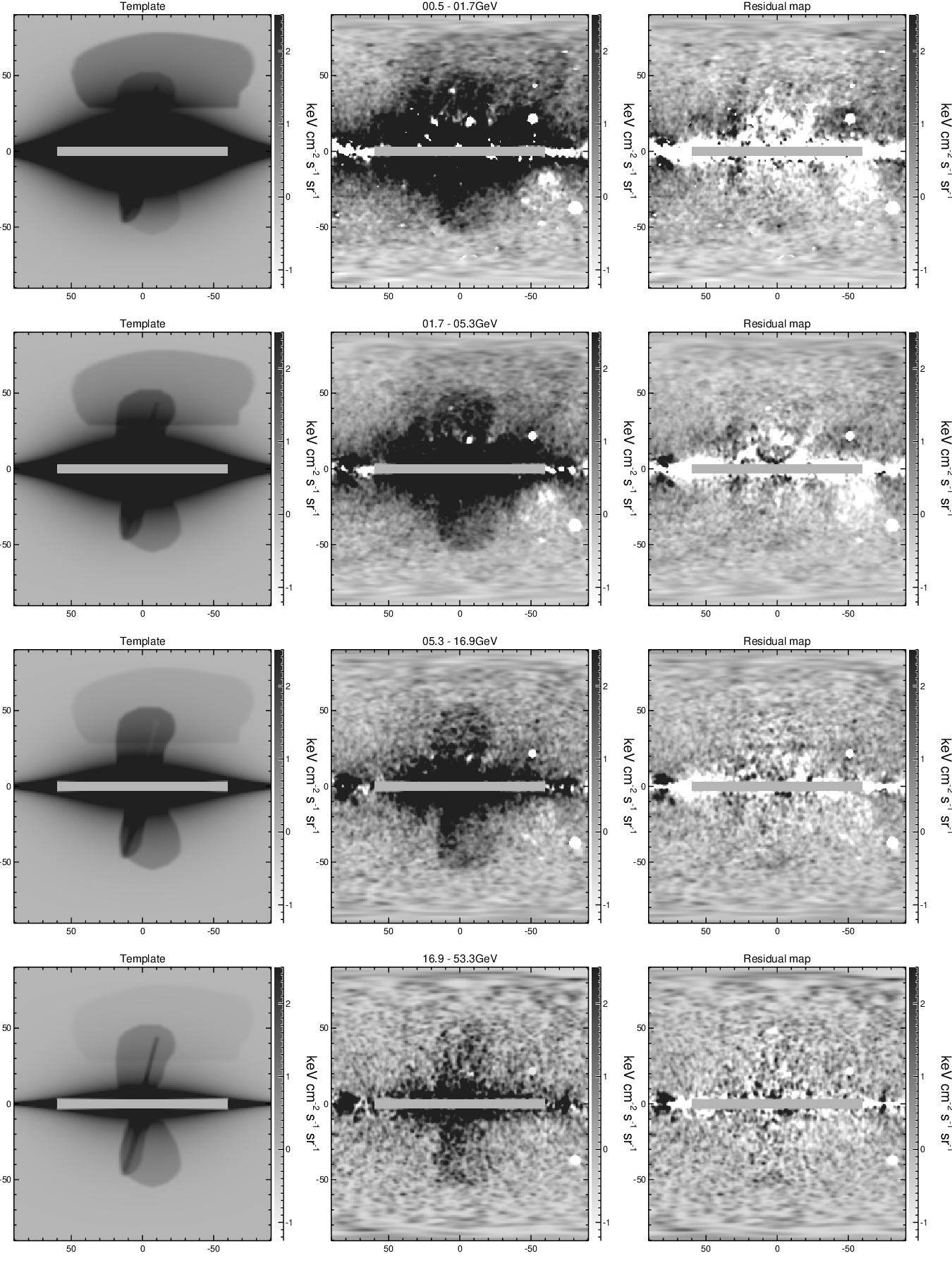}
\end{center}
\caption{Same as \reffig{cocoon}, except we add the north jet
template and split the bubble template to north and south
separately. Note that the south cocoon template is included
in the fitting. The north and south jet spectrum are
consistently hard.}
\label{fig:bubblesplitmap}
\end{figure*}


\section{Energy Spectrum of the Gamma-ray Cocoon and Jets}
\label{sec:jetspec}

We have shown in \reffig{cocoon} that the gamma-ray jets,
the cocoon, and the \Fermi\ bubbles all have a harder
spectrum than other large-scale diffuse ISM emission. In
order to measure the hardness of the spectrum and to explore
possible mechanisms responsible for the observed gamma-ray
emission, we maximize the likelihood that the maps are
described by a linear combination of spatial templates, one
for each gamma-ray emission mechanism. We fit a coefficient
to each emission component using a multi-linear regression
of simple templates, one energy bin at a time.  By combining
results from 12 logarithmically-spaced energy bands from 300
MeV to 300 GeV, we determine a spectral energy distribution for each component.  In
each fit, we maximize the Poisson likelihood of a simple
diffuse emission model involving the seven following templates:
the SFD map of Galactic dust \citep{Schlegel:1997yv}, the
simple disk model, the \Fermi\ bubbles, a template of the
\emph{Loop I} structure \citep{FermiBubble}, the gamma-ray
cocoon, the gamma-ray south jet, and a uniform background as
template to account for background gamma-ray emission and
cosmic ray contamination \citep{FermiBubble}. All maps and
templates have been smoothed by a Gaussian kernel with FWHM
of 2$\degree$ for the regression analysis. Systematic
uncertainties are dominated by the imperfect representation
of the diffuse emission by these simple templates.  We feel
that this analysis is sufficient for a rough
characterization of the cocoon and jet spectra, and provides
motivation for more thorough analysis using a more physical
model in the future.

For each set of model parameters, we compute the Poisson log
likelihood, \be \ln {\mathcal L} = \sum_i k_i\ln\mu_i -
\mu_i - \ln(k_i!), \ee where $\mu_i$ is the synthetic counts
map (i.e., linear combination of templates times exposure
map) at pixel $i$, and $k$ is the map of observed
counts. The last term is a function only of the observed
maps.  We compute errors in the Gaussian approximation by
inverting the matrix of second partial derivatives of $-\ln
{\mathcal L}$ to obtain the covariance matrix, and taking
the square root of the diagonals.  The $1\sigma$ Gaussian
error corresponds to $\Delta\ln {\mathcal L} = 1/2$.  We
refer to Appendix B of \cite{fermihaze} for more details of
the likelihood analysis. Maps of the models are constructed from
linear combinations of these templates, and the (data minus model) residual
maps in various energy bins are shown in \reffig{jetmap}. In this fit, we mask out
all pixels with Galactic latitude $|b| < 20\degree$.

So far we have described the procedure for obtaining the correlation
coefficient for each template at each energy.  However, some templates have
units (e.g., the SFD dust map is in magnitudes of $E_{B-V}$ reddening) so the
correlation coefficient has unusual units (e.g. gamma-ray emission per
magnitude).  In such a case we multiply the correlation spectrum by the
average SFD value in the ($|b| > 20\degree$) bubble region, to yield the
average spectrum of dust-correlated gamma-ray emission in this region.  
For the uniform, \emph{Loop I}, cocoon,
jet, and bubble templates, no renormalization is done.
These templates are simply ones and zeros (smoothed to the
appropriate PSF), so the south jet spectrum is simply the
spectrum of the south jet template shown in e.g. left panels
of \reffig{jetmap}, \emph{not} the mean of this
template over the whole bubble region.

The energy spectra for $\pi^0$ emission, bremsstrahlung and
IC scattering can be calculated using a sample
\texttt{GALPROP} model\footnote{\texttt{GALPROP} is a cosmic
ray propagation code. It calculates the steady
state solution to the diffusion-energy-loss equation, given
the 3D gas distribution, interstellar radiation field,
B-field model, CR diffusion assumptions, and many other
input parameters \citep{Strong:1999sv}. 
See \texttt{http://galprop.stanford.edu}} (tuned to
match locally measured protons and anti-protons as well as
locally measured electrons at $\sim20-30$ GeV), as an
indication of the expected spectral shapes. We have shown
in \citep{FermiBubble} that the energy spectra for the SFD and
the simple disk template reasonably match the model
expectations. The dust map mostly traces the $\pi^0$
emission, and the simple disk model resembles a combination
of IC and bremsstrahlung emission. The spectrum for emission
correlated with the gamma-ray jet and cocoon is clearly
significantly harder than either of these components,
consistent with a \emph{flat} spectrum in $E^2dN/dE$. This
fact coupled with the distinct spatial morphology of the
jet/cocoon system indicates that if these gamma-ray features
are generated by IC scattering of interstellar radiation
fields, then a \emph{separate} electron population must
exist in the jet/cocoon.  We also note that the spectrum of
the jet/cocoon template does not fall off significantly at
energy $\lesssim$1 GeV as the bubble spectrum does. The
fitting coefficients and corresponding errors of each
template are listed in Table 1 and 2.

The hardness of this spectrum may be the key to deciphering
the origin of the gamma-ray jet/cocoon. In~
\citep{FermiBubble}, we show that at lower energy ($E
\lesssim 1\gev$) the bubble spectrum falls sharply with
decreasing energy (becomes dramatically harder than
$-2$). With reduced statistical error from three year LAT
data and the new Pass 7 \texttt{ULTRACLEAN} event selection,
we confirm significant falling of the bubble spectrum at
lower energy but not the jet and cocoon components. This is
true whether we perform the entire analysis with
front-converting events only, or use front-converting at
$E<1$ GeV and both front- and back-converting at $E>1$ GeV
as usual.

 The
null hypothesis of zero intensity of the north (south) jet
is ruled out by 3.1$\sigma$ (4.1$\sigma$), respectively, and
5.2$\sigma$ jointly for the whole jet structure.  The same
fit simultaneously finds the cocoon with 12$\sigma$
significance.  Since the region south of the Galactic center
has less foreground emission than the north (where $\rho$ Oph is),
we focus on the south jet/cocoon for our
analysis. \reffig{spectrum} (right panel) shows the resulting energy
spectrum of the gamma-rays associated with the cocoon and
the south jet.  Like the \Fermi\ bubbles, the energy spectrum of the
jet/cocoon is harder than other diffuse
gamma-ray components, although the cocoon has a spectrum at
$< 1$ GeV different from the \Fermi\ bubbles, which suggests
its origin from a distinct population of electron CRs.  The
cocoon spectrum is consistent with $E^2 dN/dE\sim$ constant,
and the north and south jet structures have an energy
spectrum of $E^2 dN/dE \sim E^{0.2\pm 0.2}$ at latitudes
$|b| > 20\degree$. The correlation coefficients for the SFD
map and simple disk model are multiplied by the average
value of these maps in the bubble region with a $|b| >
20^\circ$ cut to obtain the associated gamma-ray
emission.  Given how hard the cocoon/jet spectra
are up to $\sim 100$ GeV, bremsstrahlung can be ruled out as
the emission mechanism based on the arguments in
 \citep{Kino:2009}.

The gamma-ray jets appear to possess north-south symmetry in
morphology without noticeable difference in intensity. To
investigate whether there is any spectral difference between
the north and south jets, we augment the 7-template fit
(dust, simple disk, \Fermi\ bubble, cocoon, south jet,
uniform, \emph{Loop I}), with a \emph{north jet}
template. We show the resulting spectrum in
\reffig{northjetspec} and the maps in \reffig{northjetmap}. We then
repeat the previous fitting procedure involving the simple
disk IC template, but splitting the bubble template into
\emph{north and south bubble} templates, and allow an
independent fit of the two jet templates along with the two
bubbles. The goal is to identify variations in the intensity
and spectral index between the northern and southern jets and
the dependence of the jet spectrum on the combination of
different emission templates.  Even given this freedom, no
significant spectral differences are found between the north
and south jets or the north and south bubbles.  There was no
significant improvement of the likelihood for such
splitting. The spectrum and the maps are shown in
\reffig{bubblesplitspec} and \reffig{bubblesplitmap},
respectively. Our conclusion is that the gamma-ray jets
appear to be north-south symmetric, both with a hard
spectrum. This statement is largely independent of our
choice of template for the disk IC emission. In summary, we
found no apparent difference between the north and the south
jets both in morphology and in energy spectrum, which
indicates they might share the same origin.


\begin{figure*}[ht]
\begin{center}
\includegraphics[width=0.8\textwidth]{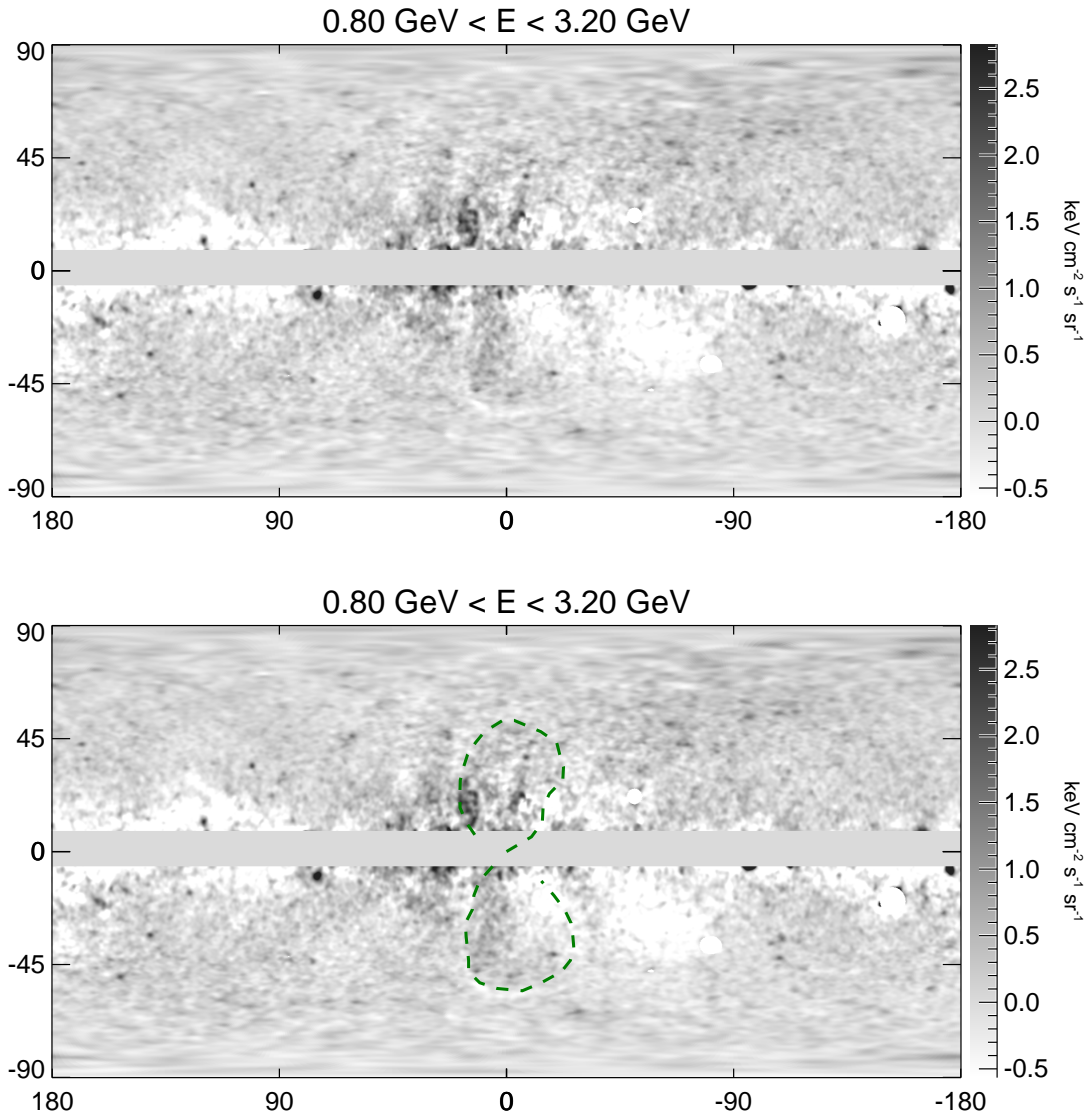}
\end{center}
\caption{Full sky
residual maps after subtracting the \Fermi\ diffuse emission
model from the \Fermi-LAT three year gamma-ray maps.
\emph{Upper panel:} point sources have been subtracted, and
large sources, including the inner Galactic disk, have been
masked. The \Fermi\ bubble structure has been included in
the diffuse emission model. We find no large-scale jet
features other than the central Galactic jet toward the
inner Galaxy. The Galactic jet feature is crossing the
Galactic center, and aligns with the long axis of the cocoon
structure. \emph{Lower panel:} the \Fermi\ bubble edge has
been marked in green dashed circles above and below the
Galactic center, overplotted on the gamma-ray map the
same as the \emph{upper panel}. Both the north and the south
jets approximately end on the edge of the Fermi bubble
structure.}
\label{fig:fullsky}
\end{figure*}


\begin{figure}[ht]
\begin{center}
\includegraphics[width=0.45\textwidth]{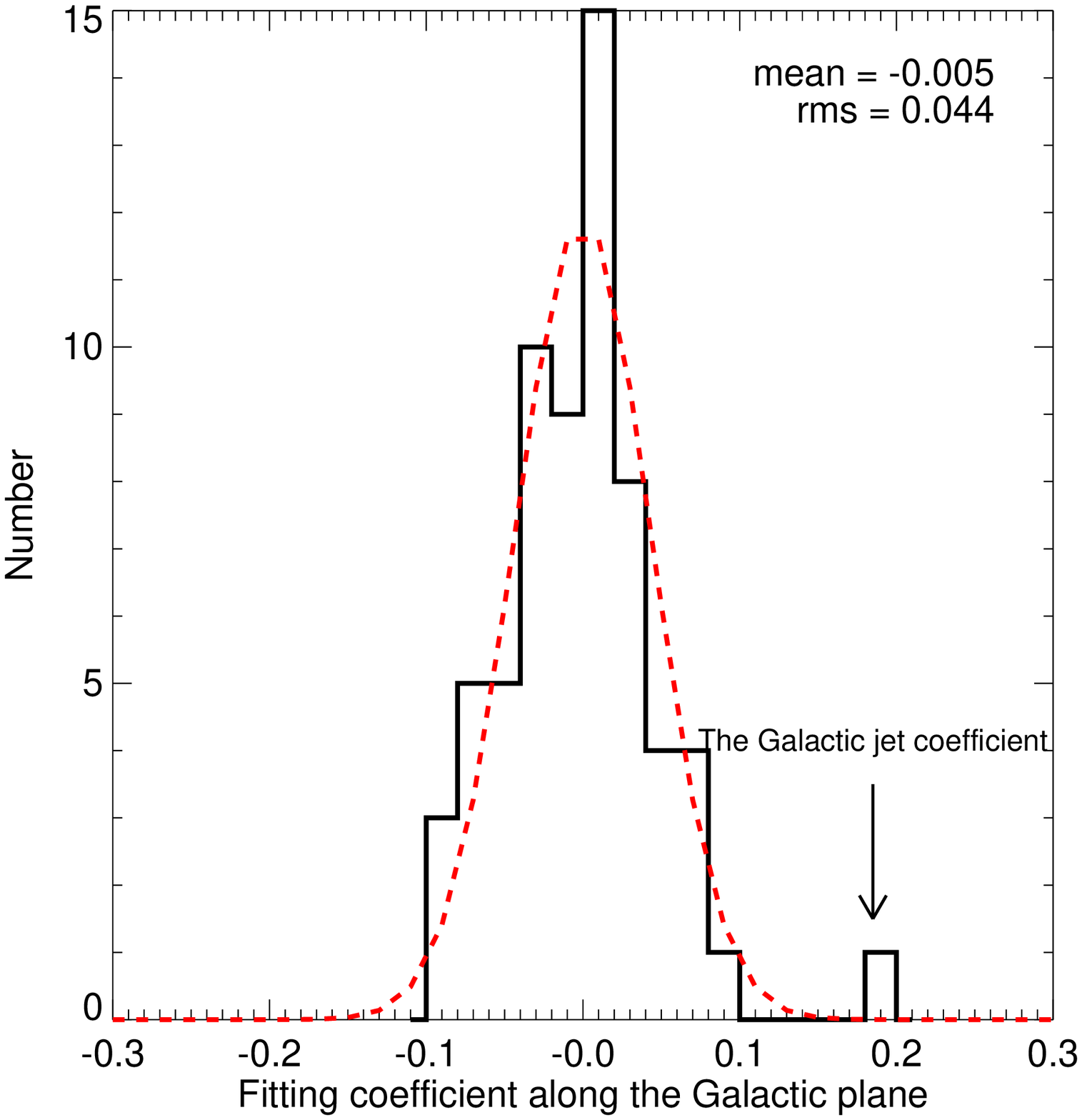}
\end{center}
\caption{As a verification of our uncertainty estimate for
  the Galactic jet structure, we have rotated the south jet
  and cocoon templates in longitude around the sky, and
  determined the best fit coefficient for each case. This
  histogram shows the distribution of the coefficients
  overplotted with the fitted Gaussian distribution. The
  coefficient distribution is well centered around zero and
  well fitted with a Gaussian noise distribution. The south
  jet template with a coefficient 0.182 is about 4$\sigma$
  significance from the background noise estimation. The
  estimated uncertainty of the jet template agrees well with
  our estimation of the significance of the south jet based
  on the Poisson likelihood fitting.}
\label{fig:jetshift}
\end{figure}

\section{Expected Radio Luminosity of the Jet}
\label{sec:radiojet}

In order to estimate the expected radio luminosity of the jet, we
must either have a physical model of the radio and gamma-ray
emission mechanisms, or a prototype object for comparison.
Very Long Baseline Array 5 GHz observations of blazars detected by LAT provide
one benchmark \citep{Linford:2011}.  This sample of
blazars shows a median 5 GHz flux of $\sim 300$ mJy and a median
gamma-ray flux ($0.1 - 100$ GeV) of $3\times10^{-8}$ photons
cm$^{-2}$s$^{-1}$, although individual objects can deviate from
this ratio by an order of magnitude or more.  BL Lac objects
have a spectrum of $dN/dE\sim E^{-1.8}$ in this energy
range, and therefore have 1/15 of their $0.1 <E< 100$ GeV
photons in the range $1-2$ GeV.  Assuming such a spectrum, a 5
GHz flux of 300 mJy corresponds to $4\times10^{-9}$ GeV
cm$^{-2}$s$^{-1}$ ($1-2$ GeV).  Therefore, the 5 GHz radio
signal corresponding to our observed $4\times10^{-7}$ GeV
cm$^{-2}$s$^{-1}$sr$^{-1}$ would be 30 Jy/sr, much smaller
than our radio limits in \reffig{radiojet}.  However,
because BL Lac objects are thought to contain a relativistic
jet pointed along the line of sight, they may be a poor
analog for the Galactic jet.

An alternative is to take a magnetic field constrained by
radio and Faraday rotation measure observations
\citep{Sun:2008} and an interstellar radiation field (taken
from \texttt{GALPROP}, 4 kpc below the Galactic center) to
compute the IC gammas and synchrotron given a power-law CR
electron spectrum.  The high-energy cutoff of the electron
spectrum is important for the IC gammas, especially for
spectra harder than $dN/dE\sim E^{-2}$. As an example, we
have computed the IC gamma-ray and synchrotron emission for
4 power laws, with spectral index in the range
$\gamma=1.5-2.5$ with a cutoff at $E_{\rm electron}=1$ TeV
(\reffig{radiojet}).  The electron spectra are normalized
such that the IC gammas go through the data points. The
synchrotron is then computed assuming a 5 $\mu$G magnetic
field and the \texttt{GALPROP} ISRF \citep{porter08} at 4 kpc below the
Galactic plane.  Upper limits from maps at 408 MHz
\citep{Haslam} and 2.3 GHz \citep{Rhodes}, and a data point
from \WMAP\ 23 GHz 4 kpc off the disk (which together serve as
an upper limit for the jet brightness \citep{Dobler:2008ww})
are included.  Most of the lines pass below the limits,
making it plausible that the radio signal corresponding to a
gamma-ray jet would not yet have been observed.  Additional
uncertainty arises from the fact that the 1 GHz radio signal
is produced by order 1 GeV electrons, whereas the gamma-ray
signal arises from order $100-1000$ GeV electrons.  If the
electron spectrum has any downward curvature over this range
in energy, the radio/microwave limits become even less
constraining.  Another uncertainty comes from the choice of
energy cutoff for the power law.  We considered spectra with
cutoffs at 3 and 5 TeV with index $\gamma\sim 3$ to fit the
gamma-ray data.  Such spectra predict a synchrotron signal
bright enough to be ruled out by the radio data.  This
implies that the required CR electron spectrum is much harder than
the locally measured index of $\gamma\sim 3-3.25$.

\section{Statistical Significance of the Jet}
\label{sec:statistics}

In \reffig{fullsky}, we show the full sky residual
maps at $0.8-3.2$ GeV after subtracting the \Fermi\ diffuse
emission model. This figure is the full sky version of
\reffig{cocoon}. We identify a pair of Galactic jet features
toward the inner Galaxy, present both above and below the
Galactic plane and aligned with the Galactic
center. There are no other significant large-scale jet
structures appearing in the full sky gamma-ray map.

Our best-fit values are derived from a maximum likelihood
analysis, in which we maximize the Poisson probability of
observing the observed counts, given a set of model
parameters.  The error bars are derived from the parameter
covariance matrix.  In our previous work
\citep{FermiBubble}, we analyzed numerous mock maps to
verify that our code gives unbiased results with the correct
uncertainties, at least in the case where the model (plus
Poisson noise) is a good description of the data.  However,
our fits contain systematic residuals, and it is necessary
to investigate to what extent these might be able to mimic a
signal.  To this end, we have repeated the 7-component fit
with the cocoon and jet structures shifted around the sky at
5 deg intervals in longitude $\ell$, and determined the
best fit coefficient for each case by fitting the same
templates as in \reffig{spectrum}. By placing the templates
at different position along the longitudes where we expect
no significant large scale diffuse structures, we can
measure the rms of the jet and cocoon coefficients, and this
rms may exceed that due to the Poisson noise.  In some parts
of the sky, (e.g. $\ell \sim 290-320\degree$), the $\pi^0$
emission is somewhat oversubtracted, which pushes the cocoon
coefficient negative, effectively renormalizing the sky
background for the jet.
As expected, the distribution of jet template coefficients
(shown in \reffig{jetshift}) is centered around zero,
consistent with a Gaussian of mean zero and
$\sigma=0.044$. The Galactic jet coefficient is more than $4\sigma$
out of the background noise distribution, which is
consistent with the estimated significance of the south jet
derived from the covariance matrix of our Poisson likelihood
fit.  We interpret this agreement to mean that systematic
errors in foreground modeling have not substantially
distorted the meaning of the south jet's $4\sigma$ formal
significance.

\begin{table*}
 \begin{center}
 \begin{tabular}{@{}rrrrrrrr}

\hline
\hline
$E$ Range (GeV) & Energy & Uniform & SFD Dust & Simple Disk  & Bubble & Bubble (1.6 yr) &  South Cocoon\\
\hline
$   0.3 -   0.5$
 &    0.4 &  1.376 $\pm$ 0.007 &  1.602 $\pm$ 0.019 &  0.451 $\pm$ 0.011 &  0.024 $\pm$ 0.020 &    0.035 $\pm$ 0.033 & 0.373 $\pm$ 0.055 \\
$   0.5 -   0.9$                                                                                                      
 &    0.7 &  1.175 $\pm$ 0.007 &  1.696 $\pm$ 0.019 &  0.458 $\pm$ 0.012 &  0.128 $\pm$ 0.021 &    0.211 $\pm$ 0.037 & 0.315 $\pm$ 0.060 \\
$   0.9 -   1.7$                                                                                                      
 &    1.3 &  0.897 $\pm$ 0.007 &  1.489 $\pm$ 0.019 &  0.383 $\pm$ 0.012 &  0.167 $\pm$ 0.022 &    0.321 $\pm$ 0.044 & 0.415 $\pm$ 0.066 \\
$   1.7 -   3.0$                                                                                                      
 &    2.2 &  0.734 $\pm$ 0.006 &  1.104 $\pm$ 0.016 &  0.324 $\pm$ 0.010 &  0.290 $\pm$ 0.018 &    0.436 $\pm$ 0.036 & 0.264 $\pm$ 0.054 \\
$   3.0 -   5.3$                                                                                                      
 &    4.0 &  0.562 $\pm$ 0.007 &  0.778 $\pm$ 0.017 &  0.249 $\pm$ 0.011 &  0.295 $\pm$ 0.021 &    0.353 $\pm$ 0.043 & 0.321 $\pm$ 0.064 \\
$   5.3 -   9.5$                                                                                                      
 &    7.1 &  0.500 $\pm$ 0.008 &  0.475 $\pm$ 0.020 &  0.170 $\pm$ 0.013 &  0.363 $\pm$ 0.025 &    0.343 $\pm$ 0.049 & 0.306 $\pm$ 0.078 \\
$   9.5 -  16.9$                                                                                                      
 &   12.7 &  0.392 $\pm$ 0.009 &  0.305 $\pm$ 0.023 &  0.117 $\pm$ 0.015 &  0.365 $\pm$ 0.030 &    0.205 $\pm$ 0.055 & 0.298 $\pm$ 0.094 \\
$  16.9 -  30.0$                                                                                                      
 &   22.5 &  0.287 $\pm$ 0.011 &  0.236 $\pm$ 0.026 &  0.111 $\pm$ 0.018 &  0.307 $\pm$ 0.035 &    0.263 $\pm$ 0.068 & 0.473 $\pm$ 0.119 \\
$  30.0 -  53.3$                                                                                                      
 &   40.0 &  0.244 $\pm$ 0.013 &  0.156 $\pm$ 0.030 &  0.028 $\pm$ 0.020 &  0.279 $\pm$ 0.041 &    0.217 $\pm$ 0.083 & 0.051 $\pm$ 0.125 \\
$  53.3 -  94.9$                                                                                                      
 &   71.1 &  0.169 $\pm$ 0.014 &  0.170 $\pm$ 0.036 &  0.014 $\pm$ 0.022 &  0.309 $\pm$ 0.050 &    0.251 $\pm$ 0.120 & 0.132 $\pm$ 0.152 \\
$  94.9 - 168.7$                                                                                                      
 &  126.5 &  0.130 $\pm$ 0.016 &  0.060 $\pm$ 0.039 &  0.014 $\pm$ 0.025 &  0.241 $\pm$ 0.057 &    0.319 $\pm$ 0.162 & 0.079 $\pm$ 0.180 \\
$ 168.7 - 300.0$                                                                                                      
 &  225.0 &  0.086 $\pm$ 0.017 &  0.045 $\pm$ 0.040 & -0.010 $\pm$ 0.027 &  0.040 $\pm$ 0.052 &   -0.015 $\pm$ 0.194 & 0.071 $\pm$ 0.173 \\
\hline
 \end{tabular}
 \end{center}
\caption{The template fitting coefficients and errors
  corresponds to \reffig{northjetspec} and
  \reffig{northjetmap}. The gamma-ray luminosity in each
  energy range is shown in the unit of $\kevflux$. For comparison, we also listed the \Fermi\ bubble luminosity from \citep{FermiBubble}. }
\label{tbl:bubblefits}
\end{table*}

 \begin{table}
 \begin{center}
 \begin{tabular}{@{}rrrrr}
\hline
\hline
$E$ Range (GeV) & Energy & North Jet & South Jet\\
\hline
$   0.3 -   0.9$
 &    0.6 &  0.235 $\pm$ 0.139 &  0.107 $\pm$ 0.112 \\
$   0.9 -   3.0$
 &    1.8 &  0.248 $\pm$ 0.163 &  0.208 $\pm$ 0.115 \\
$   3.0 -   9.5$
 &    5.6 &  0.248 $\pm$ 0.170 &  0.002 $\pm$ 0.120 \\
$   9.5 -  30.0$
 &   17.6 & --0.036 $\pm$ 0.223 &  0.318 $\pm$ 0.182 \\
$  30.0 -  94.9$
 &   55.6 &  0.526 $\pm$ 0.339 &  0.656 $\pm$ 0.266 \\
$  94.9 - 300.0$
 &  175.7 &  0.151 $\pm$ 0.446 &  0.770 $\pm$ 0.395 \\
\hline
 \end{tabular}
 \end{center}
\caption{The template-fitting Coefficients and Errors of the
  North and South jets correspond to \reffig{northjetspec}
  and \reffig{northjetmap}. The gamma-ray luminosity in each
  energy range is shown in the unit of $\kevflux$.}
\label{tbl:jetfits}
\end{table}

\section{Discussion and Conclusion}
\label{sec:conclusion}

Given the alignment of the cocoon with the southern bubble
edge, the cocoon/jet is probably associated with the bubble,
and their origins may be intertwined.  A possible scenario
in which a jet creates both the cocoon and the
\Fermi\ bubble is as follows.  The propagation of a
large-scale relativistic jet may generate a double bow-shock
structure at the head of the jet
\citep{Blandford:1974,Scheuer:1974}. Entrained energy and
matter are pushed aside due to a high pressure gradient and
create a hot cocoon around the jet. The cocoon applies
sufficiently high pressure to collimate the jet and
substantially reduce its opening angle. Continuous injection
of relativistic electrons can be produced in the forward,
reverse, and re-collimation shocks emitting synchrotron
radiation and gamma-ray IC emission. The jets deliver
kinetic energy to the surrounding interstellar gas and the
jet/medium interactions could be strong enough to accelerate
particles and produce non-thermal radiation. In this
scenario, as the jets emerge from the GC, they expand freely
until they are recollimated when their ram pressure falls to
the thermal pressure of the surrounding cocoon.  The jet
decelerates once the accumulated mass of the swept up ISM
gas becomes similar to that carried by the jet.  The cocoon
and \Fermi\ bubble could be explained by jet shocked
material (re-confinement region and cocoon) and the ambient
material could be shocked by the bow shock. However, the $15\degree$
misalignment between the jet/cocoon and the bubbles, and the
absence of an obvious northern cocoon, suggests that this is
not the whole story.  We cannot rule out the possibility
that the \Fermi\ bubbles were formed by an earlier energetic
event, while the jet might be produced more recently and be
penetrating through the older structure.




The gamma-ray jet is the first collimated jet structure
found in gamma-rays and the only collimated jet close 
enough to resolve with \Fermi-LAT. The presence of such a
large-scale Galactic jet in our Milky Way provides an ideal
nearby laboratory for studying basic questions about jet
formation, acceleration and collimation.  It is likely that 
similar systems are not uncommon.  For example, the recently
observed unusual transient source (Swift J164449.3 + 573451)
has been understood as a newly formed relativistic outflow
launched by transient accretion onto a SMBH with mass similar to
the Galactic
center SMBH \citep{Zauderer:2011,Burrows:2011}. It provides the
evidence that a normal galaxy can transform to an AGN-like
phase and produce a relativistic jet. 
Similar structures on much larger
scales have been found in galaxy clusters with radio lobes
inflated by the jets of SMBH at the center of powerful radio
galaxies \citep{McNamara:2005}.

Follow-up observations at other wavelengths are required to
advance our understanding of the jet and cocoon structures,
and their relation to the \Fermi\ bubbles.  In coming years,
\emph{eRosita} will survey the whole sky in X-rays with
approximately 30 times \emph{ROSAT} sensitivity, in the
medium energy X-ray range up to 10 keV with an unprecedented
spectral and angular resolution  \citep{eRosita}. It has the
ability to reveal important information about the state of
the material in the jet, cocoon, and bubble structures.  Such
multi-wavelength exploration will advance our understanding
not only of our own Galaxy, but of black hole accretion and
jet formation in general.

\vskip 0.15in {\bf \noindent Acknowledgments:} It is a
pleasure to acknowledge helpful conversations with Aneta
Siemiginowska, Martin Elvis, Feng Yuan, and Roman
Shcherbakov.  We acknowledge the use of public data from the
\Fermi\ data archive at
\texttt{http://fermi.gsfc.nasa.gov/ssc/}. M.S. and D.P.F. are
partially supported by the NASA Fermi Guest Investigator
Program.  This research made use of the NASA Astrophysics
Data System (ADS) and the IDL Astronomy User's Library at
Goddard (available at
\texttt{http://idlastro.gsfc.nasa.gov}).
\newpage

\bibliography{jet} 
\bibliographystyle{apj}

\end{document}